# Controllable density of states in molecular devices: the magnetic effect on cyclic molecules

Jingzhe Chen, Jin Zhang , Rushan Han*

School of Physics, Peking University, Beijing 100871, People's Republic of China

*The corresponding author. Email: rshan@pku.edu.cn

**ABSTRACT**

Through the analysis of density of states (DOS), we study two different kinds of cyclic molecules, a sodium atomic circle and a Polycyclic Aromatic Hydrocarbons (PAH) molecule respectively under an external magnetic field. The results of calculation show that original DOS peaks split into two and the positions of peaks can be modulated periodically by controlling the magnetic field. We attribute this phenomenon to the interaction between the molecular orbital magnetic momentum and the magnetic field. Generally, this effect is obscure for a usual structure but obvious for the cyclic molecules.

## INTRODUCTION

Recently, cyclic molecules have attracted more and more attention both in the biological systems and molecular electronics.[1,2] Some researchers pointed out that they were the promising candidates for the fabrication of molecular wires and molecular electronic devices.[3-5] Particularly, as a typical quantum coherent system, cyclic molecules can be used as logic gate cells.[6] To investigate the electronic property of such molecular devices, several first-principle calculation methods have been developed to simulate the corresponding situations.[7-9] In this paper, we provide a method to demonstrate the density of states (DOS) analysis of cyclic molecules under an external magnetic field.

The molecular DOS is sensitive to some external factors such as gate voltage, light and current.[10-12] As far as magnetic field is concerned, generally, the change is relatively small under an available magnetic field (~20Tesla) and there are few related calculations. Despite the fact, an observable DOS change of the specific molecule is shown under such a magnetic field and it shows clearly there exists quantum interference due to the Bohm-Aharonov phase acquired by the electron wave “propagating” on the molecule.

## METHOD

Based on density functional theory (DFT), the Kohn-Sham Hamiltonian of a molecule is written as:

$$H = \frac{1}{2m}(\vec{P} + e\vec{A})^2 + V - \vec{\mu}_s \cdot \vec{B} \qquad (1)$$

$\vec{\mu}_s$ is the spin magnetic moment, $\vec{B}$ is the magnetic vector and $V$ includes the nuclear attraction potential, the Hartree potential and the exchange potential. Here we adopt the Coulomb gauge and the magnetic vector potential $\vec{A}$ is defined as

$$\vec{A} = \frac{1}{2}\vec{B} \times \vec{r} \qquad (2)$$

Let $\vec{r} = \vec{R} + \vec{r}\,'$, here $\vec{R}$ is the coordinates of atoms, and $\vec{r}\,'$ is the relative coordinates $\vec{R}$. Assuming that the direction of the magnetic field is parallel to z axis, the Kinetic item in Hamiltonian can be written into:

$$H^{Kin} = \frac{1}{2m}(\vec{P}^2 + e\vec{B}(\vec{l}_z + \vec{L}_z) + \frac{e^2B^2}{4}(x^2 + y^2 + X^2 + Y^2 + 2xX + 2yY)) \quad (3)$$

here

$$\begin{aligned} \vec{l}_z &= (x\vec{P}_y - y\vec{P}_x) \\ \vec{L}_z &= (X\vec{P}_y - Y\vec{P}_x) \end{aligned} \qquad (4)$$

x, y, z and X,Y, Z are the components of $\vec{r}\,'$ and $\vec{R}$.

We adopt a gauge invariant atomic orbital basis: $\left|\varphi\right\rangle = cX^lY^mZ^ne^{-\alpha r^2 - i\vec{A}(\vec{R})\cdot\vec{r}}$ and $\vec{A}(\vec{R}) = \frac{1}{2}\vec{B}\times\vec{R}$. The overlap integral between the basis function is defined as:

$$\left\langle\varphi_i\middle|\varphi_j\right\rangle = \left\langle l_i, m_i, n_i \mid l_j, m_j, n_j\right\rangle \qquad (5)$$

Then we have

$$\begin{aligned} \left\langle\varphi_i \mid l_z \mid \varphi_j\right\rangle &= A_x\left\langle l_i,m_i,n_i \mid l_j,m_j+1,n_j\right\rangle - A_y\left\langle l_i,m_i,n_i \mid l_j+1,m_j,n_j\right\rangle \\ &-i(m_j\left\langle l_i,m_i,n_i \mid l_j+1,m_j-1,n_j\right\rangle - l_j\left\langle l_i,m_i,n_i \mid l_j-1,m_j+1,n_j\right\rangle) \end{aligned} \qquad (6)$$

$$\begin{aligned} \left\langle\varphi_i \mid L_z \mid \varphi_j\right\rangle &= (A_xY_j - A_yX_j)\left\langle l_i,m_i,n_i \mid l_j,m_j,n_j\right\rangle - i(X_jm_j\left\langle l_i,m_i,n_i \mid l_j,m_j-1,n_j\right\rangle \\ &-Y_jl_j\left\langle l_i,m_i,n_i \mid l_j-1,m_j,n_j\right\rangle - 2X_j\alpha_j\left\langle l_i,m_i,n_i \mid l_j,m_j+1,n_j\right\rangle + 2Y_j\alpha_j\left\langle l_i,m_i,n_i \mid l_j+1,m_j,n\right\rangle) \end{aligned} \qquad (7)$$

$$\left\langle\varphi_i \mid x^2 \mid \varphi_j\right\rangle = \left\langle l_i,m_i,n_i \mid l_j+2,m_j,n_j\right\rangle \qquad (8)$$

$$\left\langle\varphi_i \mid xX \mid \varphi_j\right\rangle = X\left\langle l_i,m_i,n_i \mid l_j+1,m_j,n_j\right\rangle \qquad (9)$$

London approximation is used to calculate the integral,[13,14]

$$\left\langle\varphi_i\middle|\varphi_j\right\rangle = \left\langle\phi_i\middle|\phi_j\right\rangle e^{iL_{ij}} \qquad (10)$$

here, $\left|\phi\right\rangle = cX^lY^mZ^ne^{-\alpha r^2}$ , $L_{i,j} = \frac{1}{2}(\vec{A}_j - \vec{A}_i)\cdot(\vec{R}_i + \vec{R}_j)$.

As the other part of Hamiltonian, $V$ is calculated with Gaussian package.

Then the retarded Green function matrix can be calculated with:

$$G^R(E) = ((E+\Gamma)S - F)^{-1} \qquad (11)$$

$F$ is the matrix representation of the Hamiltonian operator on the basis $\left|\varphi\right\rangle$,and $S$ is the overlap matrix of the basis. For a single molecule, the energy levels are discrete, $\Gamma$ is a broadening function which can be set as a pure imaginary constant approximately (in this paper, $|\Gamma|$ is set as 0.01eV).

The DOS is calculated with the formula:

$$DOS(E) = -\frac{1}{\pi}\mathrm{Im}(Tr(G^R(E)S)) \qquad (12)$$

## SAMPLES AND RESULTS

In this paper we will discuss two samples: a sodium atomic circle and a Polycyclic Aromatic Hydrocarbons (PAH) molecule (see Figure 1), because electrons in the sodium circle are nearly free electrons in a weak periodic potential and PAH is a typical organic cyclic molecule stable in nature. The external magnetic field is applied on them perpendicularly to the circle plane.

Early in 1990s, an Iron atomic circle has been successfully made by STM on Copper (Cu) surface (111), which proves an atomic circle is possible in experiments[15,16]. Here we adopt a sample with 80 sodium atoms, and the bond length is 3.3Å. We choose BLYP as the functional method and LANL1MB as the basis. Figure 2 shows the DOS of a sodium atomic circle within a wide energy scope.

In Figure 2 we can see the DOS of a sodium atomic circle is a series of discrete peaks near the Fermi level, here the Fermi level is approximated by the average of the highest occupied level (HOMO) and the lowest unoccupied level (LUMO). The DOS outline is similar to the result of an infinite long sodium chain (the bond is 3.3 Å as well), and they both have two peaks near -4.0eV and -1.5eV. The two peaks are led by one-dimensional Von Hoff singular for the 's' electrons in the sodium atomic circle are nearly free electrons. However, for the former there are several discrete peaks between the two peaks. It can easily be understood that 's' electrons are constrained in a finite periodic lattice field where we can use the Born-von Karman boundary condition. Here the wave vector is $k = \frac{l}{Na}(2\pi)$ , $N$ is the number of atoms, $a$ is the lattice constant, and $l$ is an integer satisfying $-N \le l \le N$ .Using zero-order approximation, the corresponding energy is $E = \frac{\hbar^2 k^2}{2m} + V$ ($V$ is the average field) and the spectrum of molecule is a series of discrete levels: $E = \frac{2\pi^2\hbar^2 l^2}{m(Na)^2} + V$ .

Figure 3 shows the DOS under different magnetic field. We focus on the ‘s’ electrons region below the Fermi level. As we can see, there are five DOS peaks in this energy scope for a zero field with the period $E \approx 0.1eV$ . While the flux increases, each original DOS peak splits into two peaks moving far away from each other gradually, and the total number of peaks is switched to ten. When the flux reaches a specific value $0.5\Phi_0$ ( $\Phi_0 = h/e$ is the quantum flux), those peaks combine together two by two, then the total number of DOS peaks returns to five and the period keeps the constant.

We also notice at the Fermi level the DOS change is very great and nearly like an impulse function, which we can use to modulate the DOS there precisely. It is the basis of the control of logic state for a molecular device.

**PHYSICAL MODEL**

We explain the phenomena with a physical model. In such structures, there exist two degenerate Blöch wave vectors with the same energy, which are marked as $K_1$ and $K_2$ in Figure 4, and one of them is clockwise and the other is counterclockwise. Each Blöch wave vector satisfies $K \cdot L = 2n\pi$ , $L$ is the normal vector of the circle with its modulus circumference. When there is an external magnetic field, the wave vector should satisfy the equation $K \cdot L + \frac{e}{\hbar} A \cdot L = 2n\pi$ .

Then we have $k_1 = \frac{2n\pi}{l} - \frac{e}{\hbar}|A|$ , and $k_2 = \frac{2n\pi}{l} + \frac{e}{\hbar}|A|$ , $k_1$ and $k_2$ are the modulus of $K_1$ , $K_2$ respectively, $l$ is the mod of $L$ .So the degeneration is released.

We use the zero order approximation to estimate the energy $E = \frac{\hbar^2 k^2}{2m} + V$ . Then we get the energy difference after applying the magnetic field, here $\Delta E = \frac{eB}{2m} n\hbar$ if we neglect the quadratic item of $A$. To be clear, we define the molecular orbital magnetic momentum as $\mu_{MO} = \frac{en}{2m}\hbar$ , $-N \le n \le N$ .The total Hamiltonian can be written as: $H = \frac{1}{2m} P^2 + V + \mu_{MO} B$ .

In Figure 3, when the degeneration recovers, it should satisfy the energy condition $\frac{\hbar^2}{2m}\left(\frac{2n\pi}{l}\right)^2 - \frac{ne}{2m}\hbar B = \frac{\hbar^2}{2m}\left(\frac{2(n-1)\pi}{l}\right)^2 + \frac{(n-1)e}{2m}\hbar B$ or the wave vector condition $\frac{2n\pi}{l} - \frac{e}{\hbar}|A| = \frac{2(n-1)\pi}{l} + \frac{e}{\hbar}|A|$ . Then we can substitute $!A| = \frac{Bl}{4\pi}$ into $\frac{e}{\hbar}|A| = \frac{\pi}{l}$ and get $\Phi = \frac{Bl^2}{4\pi^2} = \frac{h}{2e} = 0.5\Phi_0$ , which agrees well with the result of our calculation.

**COMPARISION AND DISCUSSION**

Let's take a look another molecule named PAH, which also has a closed circular shape[17]. There are totally 30 condensed aromatic rings in this molecule and the C-C bond length is set as 1.414Å. The functional method BLYP and basis LANL1MB are also adopted.

The DOS behavior of PAH molecule is different from that of sodium atomic circle with two points. Firstly，due to specific sample structure the DOS peaks of PAH molecule are not periodic (see Figure 5).The second, we can also see that four DOS peaks below the Fermi level split under a relative small magnetic field while the fifth not and the reason is explained as below.

We conclude the splitting of DOS is also influenced by the shapes of molecular orbitals. In a sodium atomic chain, all the electrons are constrained in the narrow potential and they have only one path when traveling around the molecule, while there are much more paths in the PAH molecule. When the fluxes through those paths are different, each path will introduce an additional phase, and as an average result, the DOS splitting effect will be smeared out. It is a broadening due to averaging paths (BAP). Figure 6 shows four of the molecular orbital shapes in the PAH molecule. The subfigures a1 and a2 show the shapes of two generate molecular orbitals corresponding to the fourth DOS peak below Fermi level (E=-3.51eV), we can see from them that the densities of the outer circles are more than those of the inter circles, while for the molecular orbitals corresponding to the fifth DOS peak (E=-3.56eV) below Fermi level (see the subfigures b1 and b2), the densities of the outer circles and the inner circles are equal. We suppose the difference of probability distribution can explain the phenomenon in Figure 5 that the former DOS peak with a narrower BAP splits and the latter DOS peak with a wider BAP does not when the flux is $0.16\Phi_0$ (The flux is calculated with the inner circle area).

We consider two extreme situations to estimate the BAP of PAH molecule. One is that the electron density is concentrated in the inner loop and we have $\Delta E_1 = \frac{e\hbar}{2m}\frac{\Phi}{S}$, the other in the outer loop and we have $\Delta E_2 = \frac{e\hbar}{2m}\frac{\Phi}{S+\Delta S}$. (S is the area of inner circle) And the BAP is the difference between them which is $\frac{en\hbar\Phi}{2mS}\Delta S$, nearly 0.015eV in the sample when $\Phi = 0.16\Phi_0$. We can see the BAP is linear with $\frac{\Delta S}{S}$, so for a usual molecule, the BAP will be very big and the splitting of DOS will be obscure.

A feasible DOS-controllable molecular device can be obtained with a PAH molecule including nearly 60 condensed aromatic rings and the magnetic field needed to modulate the DOS is about 0~20 Tesla. We also know form the inference above, the BAP of the PAH molecule including 60 rings will be 0.004eV, much smaller than that of the PAH molecule including 30 rings, so the effect of DOS splitting will be so obvious that can be used in practice.

In conclusion, we study the change of the molecular DOS under an external magnetic field. Our calculation results show that some DOS peaks will split into two for cyclic molecules and it provide a way to control the molecular DOS precisely. We explain the effect with a physical model, which agree well with our calculation. We also discuss what molecules are feasible for a DOS-controllable molecular device.

**ACKNOWLEDGEMENT**

The work is supported by the National Natural Science Foundation of China (NO. 90207009, 90206048, 90406014).

**FIGURE CAPTIONS**

Figure 1(Color online) Two molecules for DOS calculation. The left is a sodium atom circle, with 80 atoms in total and the bond length 3.3Å; the right is a PAH molecule, including 30 condensed aromatic rings.

Figure 2 (Color on line) DOS of a sodium atom circle including 80 atoms. The inset is DOS of an infinite long sodium chain satisfying periodical boundary condition (PBC).

Figure 3 (Color online) DOS of a sodium atomic circle under different magnetic fields.

Figure 4 Diagram of theoretical model.

Figure 5 (Color online) DOS of PAH molecular.

Figure 6(Color online) The orbital shapes of four molecular orbitals in the PAH

molecule. a1, a2 are the two generate molecular orbitals corresponding to the fourth DOS peak below Fermi level (E=-3.51eV), and b1, b2 are the two generate molecular orbitals corresponding to the fifth DOS peak below Fermi energy (E=-3.56eV).

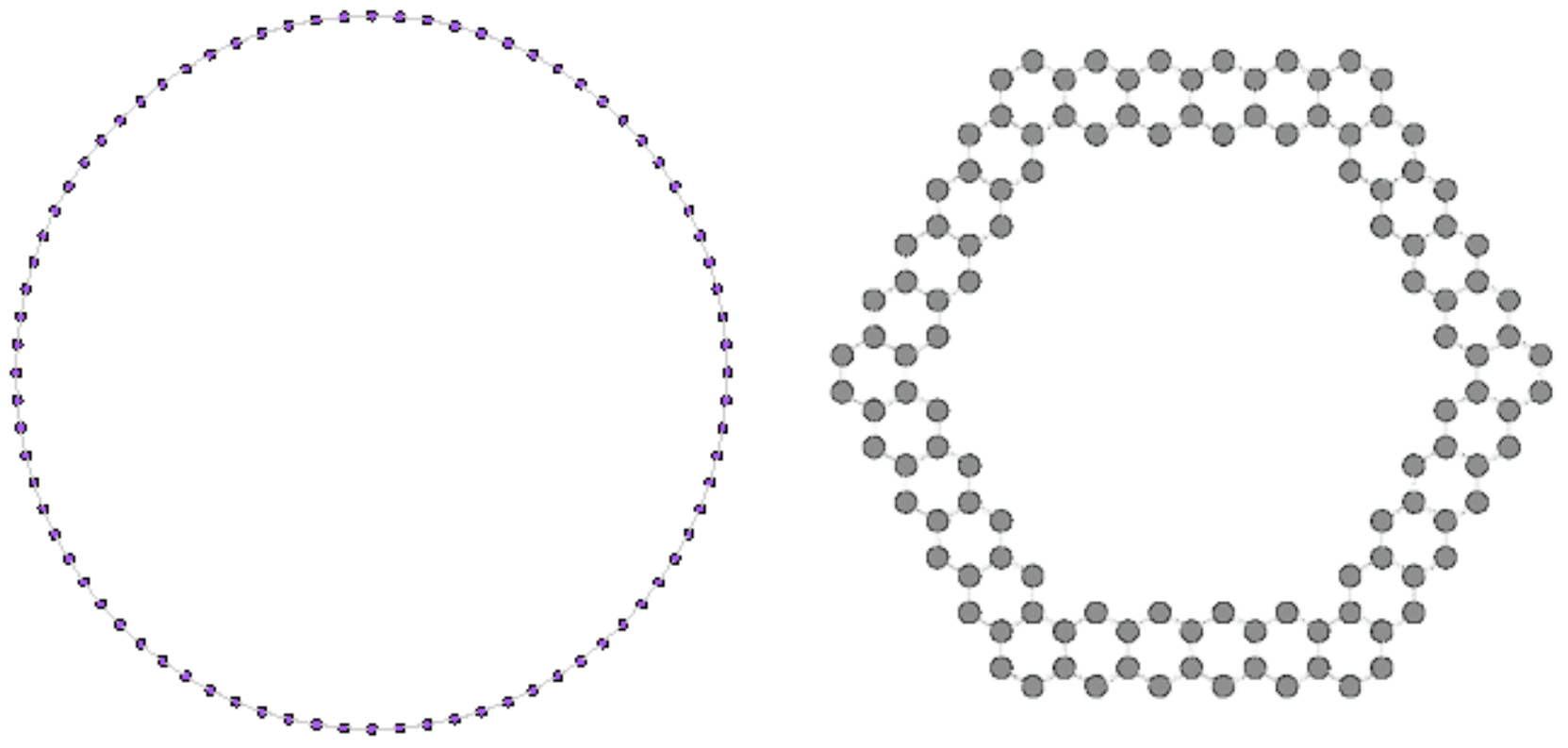

Figure 1: submitted to Physical Review B

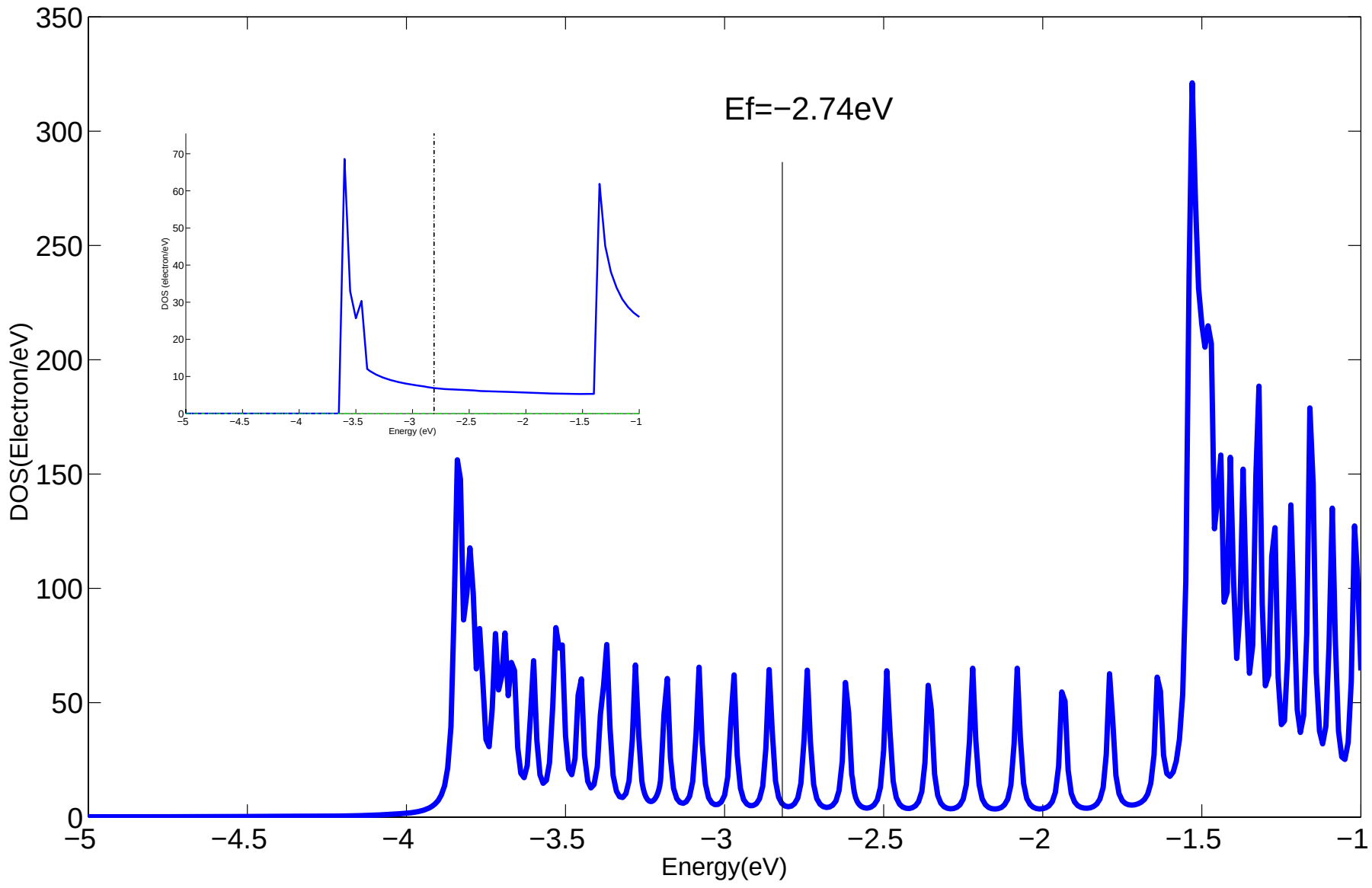

Figure 2: submitted to Physical Review B

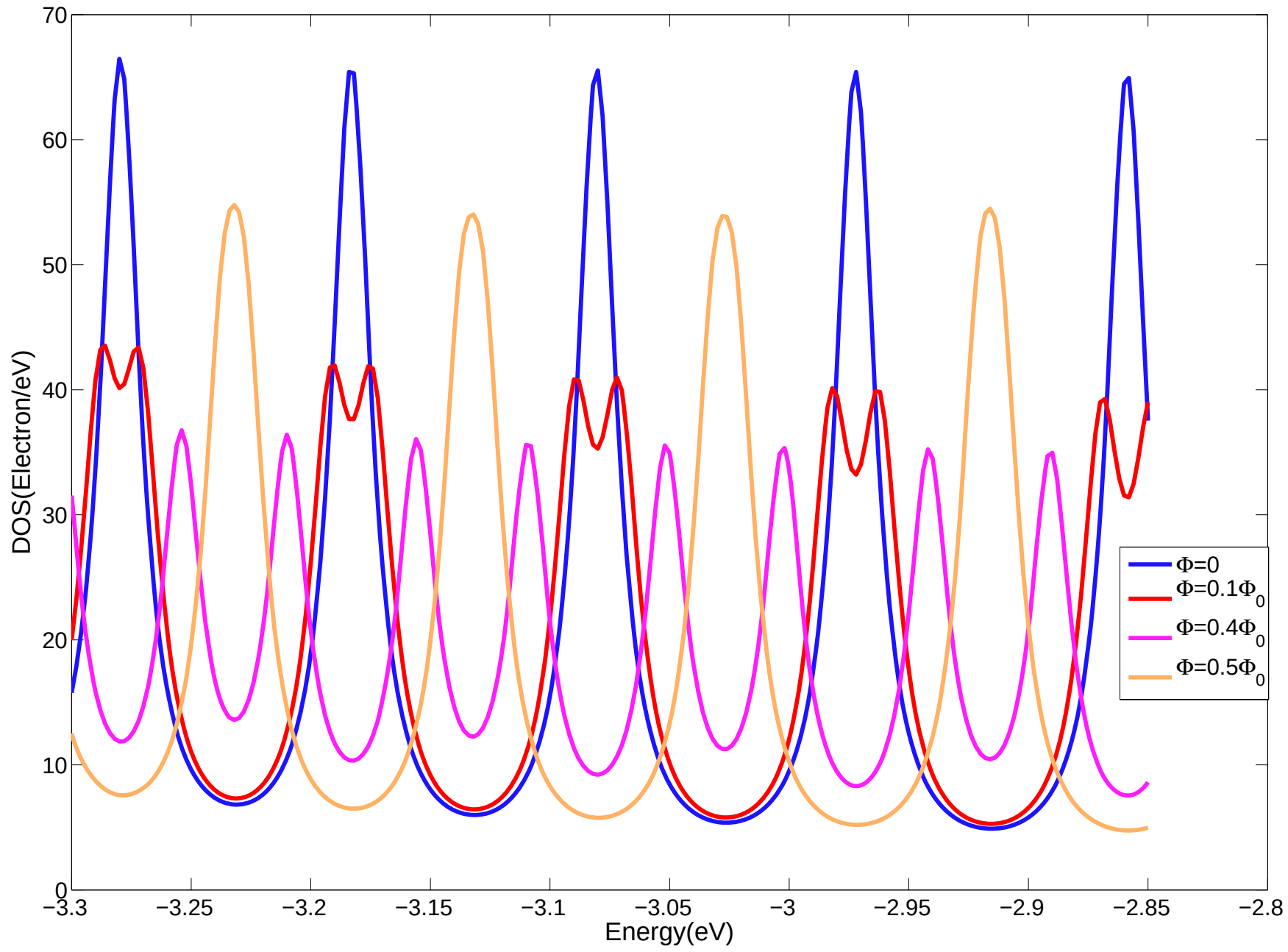

Figure 3: submitted to Physical Review B

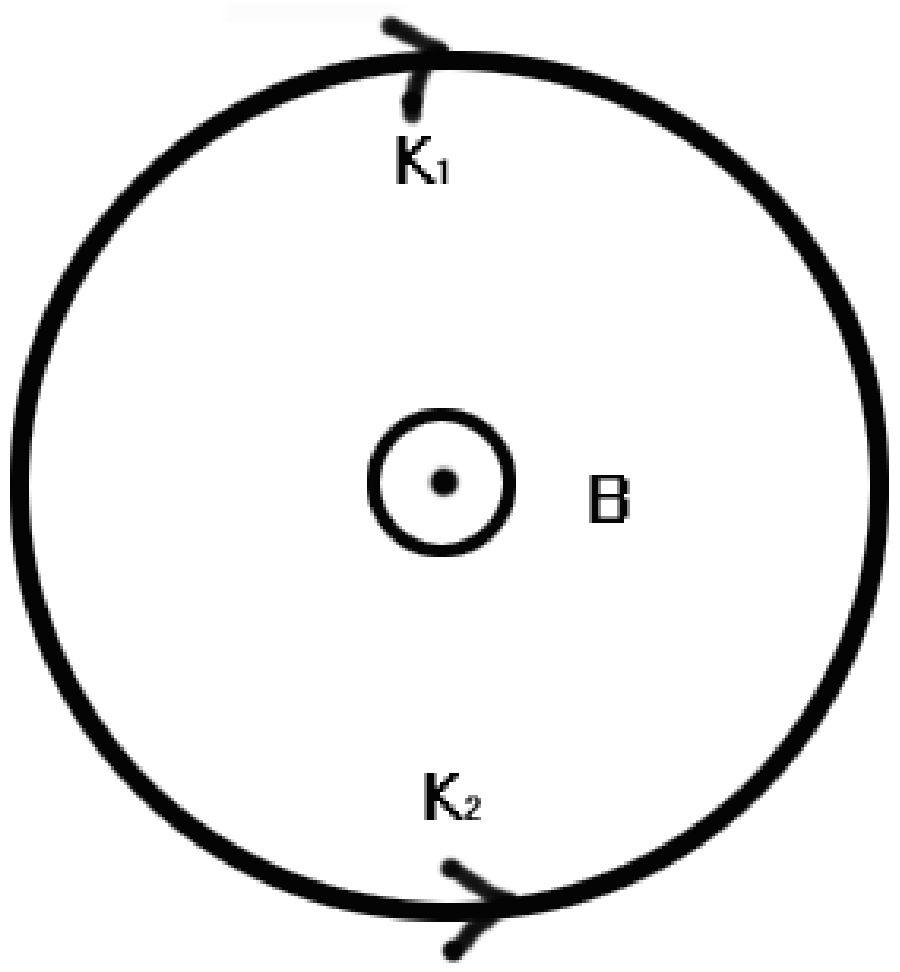

Figure 4: submitted to Physical Review B

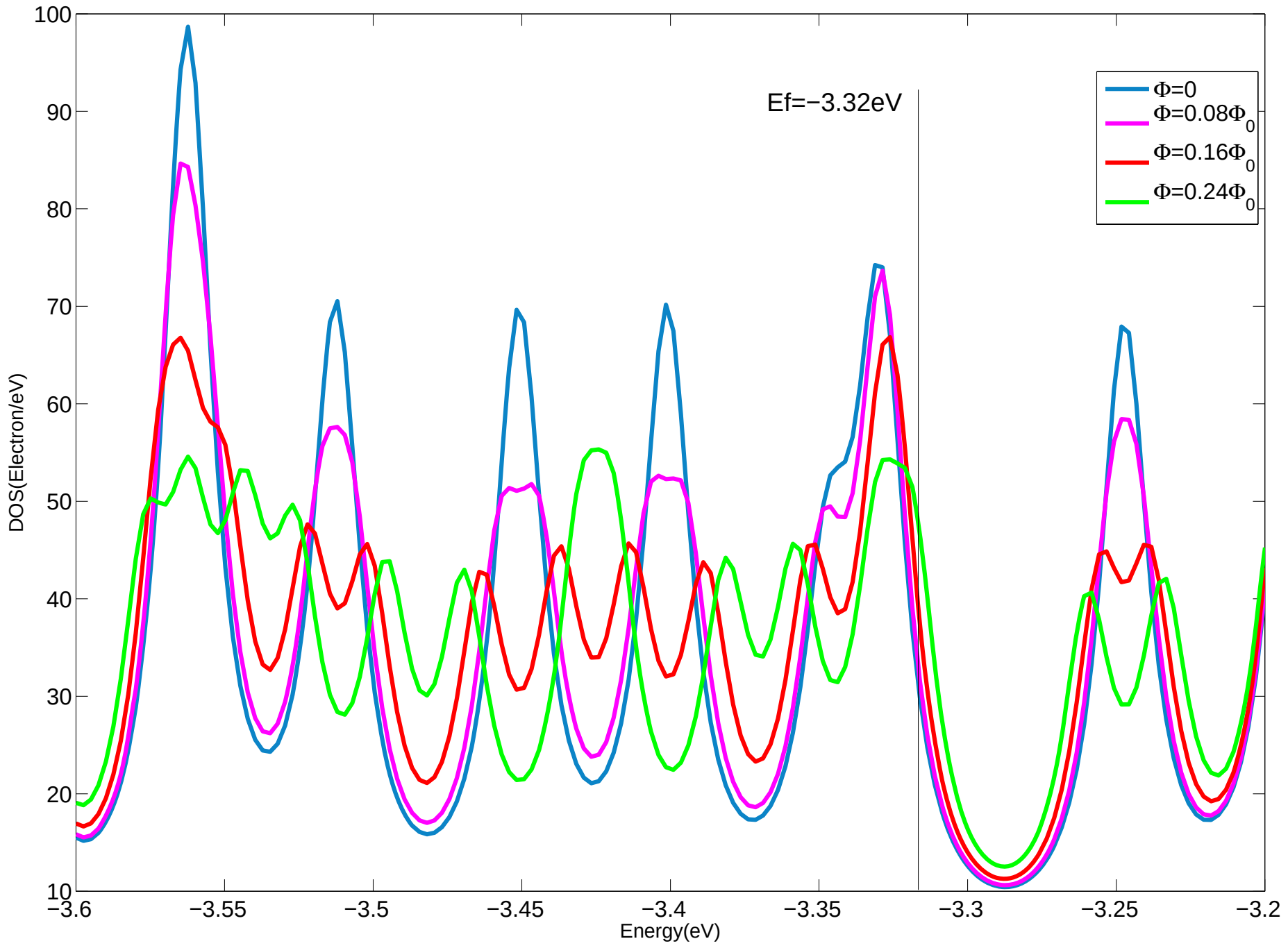

Figure 5: submitted to Physical Review B

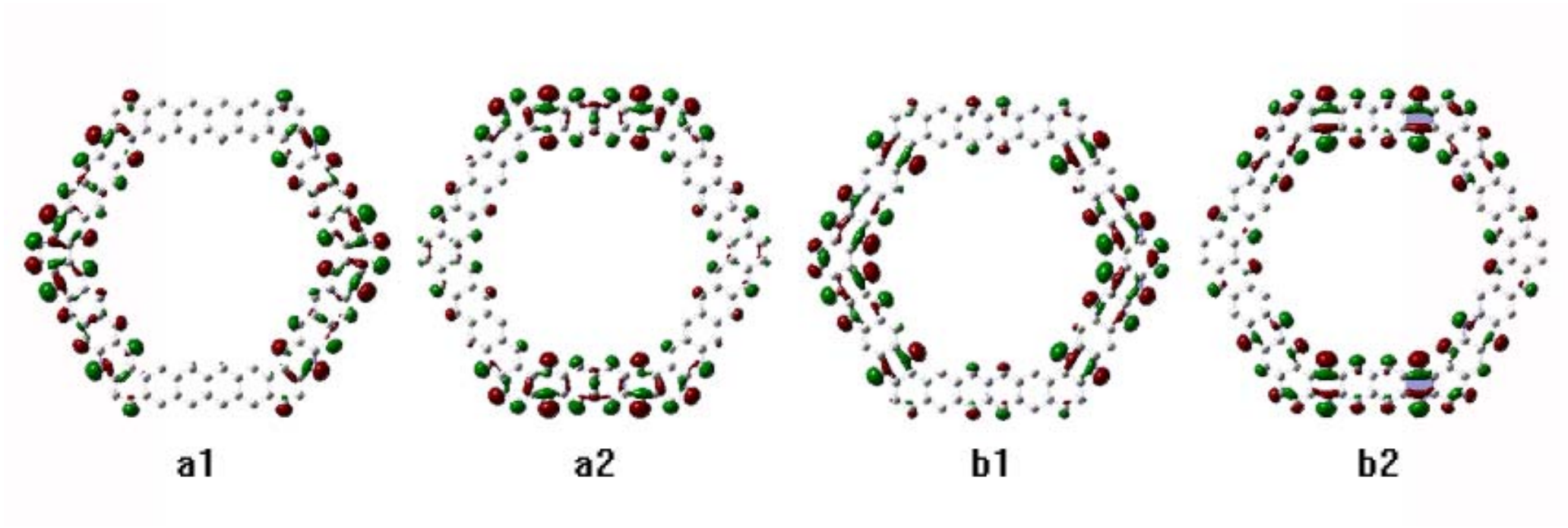

Figure 6: submitted to Physical Review B